\documentclass[superscriptaddress,twocolumn,aps,prb,reprint]{revtex4-1}
\usepackage{graphicx}
\usepackage{epstopdf}
\usepackage{braket}
\usepackage{mathrsfs}
\newcommand{\nn}{\text{Ni}_4}
\newcommand{\zz}{\text{Zn}_4}

\newcommand{\Do}{\Delta_{\pm2}^{(1)}=4.64(2)}
\newcommand{\Dt}{\Delta_{\pm2}^{(2)}=3.72(20)}
\begin{document}

\title{Precision ESR Measurements of Transverse Anisotropy in the Single-molecule Magnet Ni$_4$}

\author{Charles A. Collett}
\affiliation{Department of Physics and Astronomy, Amherst College, Amherst 01002, USA}
\author{Rafael A. All\~ao Cassaro}
\affiliation{Instituto de Qu\'imica, Universidade Federal do Rio de Janeiro, Rio de Janeiro, RJ, 21941-909 Brazil}
\author{Jonathan R. Friedman}
\email[Corresponding author. ]{jrfriedman@amherst.edu}
\affiliation{Department of Physics and Astronomy, Amherst College, Amherst 01002, USA}
\date{\today}

\begin{abstract}
We present a method for precisely measuring the tunnel splitting in single-molecule magnets using electron-spin resonance, and use these measurements to precisely and independently determine the underlying transverse anisotropy parameter, given a certain class of transitions. By diluting samples of the SMM $\nn$ via co-crystallization in a diamagnetic isostructural analogue we obtain markedly narrower resonance peaks than are observed in undiluted samples. Using custom loop-gap resonators we measure the transitions at several frequencies, allowing a precise determination of the tunnel splitting. Because the transition under investigation occurs at zero field, and arises due to a first-order perturbation from the transverse anisotropy, we can determine the magnitude of this anisotropy independent of any other Hamiltonian parameters. This method can be applied to other SMMs with tunnel splittings arising from first-order transverse anisotropy perturbations.
\end{abstract}

\maketitle

\begin{figure}[bth]
\includegraphics[width=1\linewidth]{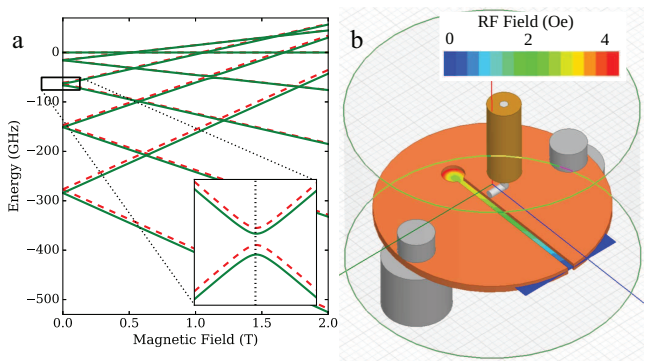}
\caption{\label{fig:lgrenfig}(a) Energy-level diagram for $\nn$ as a function of field along the z axis. Inset: Zoomed-in view of the avoided crossing for the $\ket{\pm2}$ states near zero field. Solid green and red dashed lines show the levels corresponding to the two different conformational states of the molecule. (b) Schematic of LGR inside a copper shield, with excitations provided by an antenna. The color map in the central loop and gap shows the magnitude of the rf magnetic field for a given excitation amplitude, yielding a strong uniform field in the center of the loop. Produced using ANSYS Electromagnetics Suite, Release 16.2.}
\end{figure}

Single-molecule magnets (SMMs) are spin systems ($S>\frac{1}{2}$) with an energy barrier separating different spin-orientation states. They exhibit many interesting phenomena such as magnetization tunneling \cite{friedman_macroscopic_1996-1,friedman_single-molecule_2010} and geometric-phase interference of tunneling paths.\cite{wernsdorfer_quantum_1999-1} Many of their properties can be tuned through chemical engineering and, as such, they have the potential to be exploited as qubits.  SMMs will typically crystallize with $\gtrsim10^{15}$ molecules in a crystal.  Intramolecular magnetic interactions can be strong, leading to a rigid spin-$S$ ground state. However, the molecules are well separated in the crystal lattice, making intermolecular exchange interactions between them negligible; dipole interactions are weak enough that, in the $\gtrsim1~$K temperature range, the system behaves as a paramagnet. The low-energy dynamics of most SMMs are well described by an effective ``giant spin" Hamiltonian:
\begin{equation}\label{eq:spinham}
\mathscr{H}=-DS_{z}^{2}-AS_{z}^{4}+g_z \mu_B B_z S_z+\mathscr{H^\prime},
\end{equation}
where $D$ and $A$ are axial anisotropy parameters, $g_z$ is a g factor, and $B_z$ is the applied magnetic-field component along the z axis.  $\mathscr{H^\prime}$ contains terms that do not commute with $S_z$.  With $D>0$ and $A>0$, this Hamiltonian describes a system in which the spin has lowest energy when parallel or antiparallel with z, the easy axis.  In the absence of $\mathscr{H^\prime}$, $S_z$ is a conserved quantity and the levels can be identified with values of the magnetic quantum number $m$.  Figure \ref{fig:lgrenfig}(a) illustrates the dependence of these levels of $B_z$.  $\mathscr{H^\prime}$, which contains transverse anisotropy terms and, perhaps, transverse field components breaks the symmetry of the molecule.  These terms permit tunneling between levels.  Near where the field brings different $m$ states close together, an avoided level crossing occurs (inset of Fig.~\ref{fig:lgrenfig}(a)), producing a so-called  ``tunnel splitting" -- the minimum energy gap between the two levels.  In this paper, we will focus on an SMM with four-fold symmetry, for which Eq.~$\mathscr{H^\prime}$ is given by
\begin{equation}\label{eq:hprime}
\mathscr{H^\prime}=C\left(S_{+}^{4}+S_{-}^{4}\right).
\end{equation}

The anisotropy parameters in the Hamiltonian for an SMM are often determined through electron-spin resonance (ESR) spectroscopy.  Such experiments are often done at high frequencies/high fields, where the Zeeman energy dominates the spectrum.  Parameters are extracted by fitting the resulting spectra with predictions based on the Hamiltonian.\cite{hill_high-sensitivity_1998,barra_high-frequency_2000,edwards_high-frequency_2003,hill_definitive_2003,hendrickson_origin_2005,kirman_origin_2005,barra_origin_2007,lawrence_disorder_2008,ghosh_multi-frequency_2012,sorace_origin_2013} This approach yields values for axial anisotropy terms with high precision.  To measure transverse anisotropy terms, one typically needs to apply the field in the spin's hard (x-y) plane, and analysis requires a multiparameter fit involving both axial and transverse anisotropy constants.  Because the transverse terms are generally significantly smaller than the axial anisotropy terms, they can typically only be determined to little more than one digit of precision.  Tunnel splittings, when sufficiently small, can also sometimes be inferred from dynamical magnetization measurements using a Landau-Zener technique.\cite{wernsdorfer_landau-zener_2000}  In this work, we present a low-frequency ESR method to directly measure the zero-field tunnel splitting in SMMs with splittings of order $\Delta\approx1-10$~GHz. By working with dilute orientationally ordered crystals and custom-designed resonators we precisely measure a tunnel splitting that is determined through first-order perturbation theory.  This allows us to establish the transverse anisotropy independently from any other Hamiltonian parameters.

Although dipole interactions between molecules in a crystal are weak, they are sufficient to cause significant inhomogeneous broadening of ESR lines and give rise to decoherence of the quantum spin state, diminishing the efficacy of these systems as qubits. One method for reducing dipolar interactions in molecular-spin systems is applying high fields to polarize the system and, thereby, reduce   decoherence.\cite{takahashi_coherent_2009,takahashi_decoherence_2011} Another approach is to dilute the sample, spacing the magnetic molecules apart within a diamagnetic environment either by dissolving samples in an appropriate solvent\cite{ardavan_will_2007,schlegel_direct_2008} or by co-crystallizing molecules with diamagnetic analogues.\cite{henderson_control_2008,vergnani_magnetic_2012,moro_coherent_2013,repolles_spin-lattice_2014,Shiddiq_enhancing_2016} The co-crystallization technique has the advantage that the system is crystalline and the molecules therefore retain orientational order. In addition, recent experiments have focused on atomic-clock transitions as another method of minimizing decoherence: by working at an avoided level crossing, the decoherence time $T_2$ can be significantly enhanced.\cite{Vion_manipulating_2002,Wolfowicz_atomic_2013,Shiddiq_enhancing_2016}
Our method allows us to precisely determine the tunnel splitting and thereby permits pulsed experiments to be tuned precisely to the clock transition, maximizing spin coherence.

Direct measurements of an SMM tunnel splitting have recently been done by Shiddiq \textit{et al.}\cite{Shiddiq_enhancing_2016} They studied the SMM HoW$_{10}$, diluted by co-crystallization, and measured the splitting $\Delta_{\pm4}$ between $m=\pm4$ spin states. In that case, where $\Delta m=8$, the states are coupled through second-order perturbation theory in $\mathscr{H^\prime}$, meaning that the splitting depends on both $C$ and $D$: $\Delta_{\pm4}\propto C^2/|D|$. By performing a similar experiment on a system with states connected through a first-order perturbation, we are able to directly measure $C$ for this system with no reliance on other Hamiltonian parameters.

We investigated the SMM [Ni(hmp)(dmb)Cl]$_4$ (``$\nn$"), a system with spin $S=4$,\cite{yang_fast_2006} where hmp stands for 2-hydroxymethylpyridine and dmb for 3,3-dimethyl-1-butanol. We studied $\nn$ diluted by co-crystallization with $\zz$, an isostructural diamagnetic molecule. $\nn$ was synthesized using a previously reported procedure.\cite{yang_fast_2006} The [Zn(hmp)(dmb)Cl]$_4$ (``$\zz$") complex was synthesized using the same experimental procedure, but replacing NiCl$_2\cdot$4H$_2$O by an equimolar amount of ZnCl$_2\cdot$4H$_2$O.  To prepare dilute crystals of 5\% $\nn$, 1.75 mg of $\nn$ and 35 mg of $\zz$ were dissolved in a mixture of 0.25 g of dmb and 1.7 mL of dichloromethane. Light green crystals were obtained by slow evaporation of the solvent at room temperature. The unit-cell parameters and face indexes of all compounds were determined with a Bruker Venture diffractometer using graphite and monochromatic Mo K$_\alpha$ radiation ($\lambda$ = 0.71073 \AA) at room temperature. These parameters were found to be in good agreement with the parameters previously reported for $\nn$. Metal analysis was performed by dissolving a sample in concentrated nitric acid; determination of the amount of zinc and nickel was performed using a Varian AA240FS atomic absorption spectrometer,  confirming the expected  5:95 ratio of Ni:Zn in the crystal.   High-frequency ESR spectra on 5\% $\nn$ are similar to spectra from non-dilute $\nn$, indicating that dilution leaves many of the $\nn$ molecules intact.  Some small peaks that appear only in the spectra for the dilute sample presumably represent the result of ion exchange during crystallization, producing some molecules in the crystal that are NiZn$_3$ or Ni$_3$Zn.  These ``contamination peaks" are not observable at the low frequencies of the present study.  The results presented below relate to spectral features that can unambiguously be associated with intact $\nn$ molecules.

$\nn$ can be well described by Eqs.~\ref{eq:spinham} and \ref{eq:hprime}, and has a significant transverse anisotropy term $C$, which produces a tunnel splitting of $\Delta_{\pm2}\sim4$~GHz between the $m=\pm2$ states at zero field. Figure \ref{fig:lgrenfig}(a) shows the energy-level diagram for $\nn$, with the inset highlighting this splitting. At low temperatures, $\nn$ undergoes a transition into two distinct ligand conformational states (isomers), each of which have slightly different energies as shown by the green solid and red dashed lines in Fig.~\ref{fig:lgrenfig}(a), leading to a doubling of the ESR spectra.\cite{lawrence_disorder_2008,chen_observation_2016} The effects of this conformational change on the spin Hamiltonian are not fully understood, allowing the possibility that the four-fold symmetry of $\nn$ is broken, which could introduce a second-order transverse anisotropy term. In the absence of evidence for such symmetry-breaking, or higher-order transverse anisotropy terms such as $S_z^2\left(S_{+}^{4}+S_{-}^{4}\right)+h.c.$, we assume that the only significant contribution to the splitting is the fourth-order ``$C$ term".

For fields applied along the easy axis, the field dependence of the level splitting near an avoided crossing [cf. the inset to Fig.~\ref{fig:lgrenfig}(a)] can be well described by
\begin{equation}
f_{m,m^\prime}=\sqrt{\Delta_{m,m^\prime}^2+\left[g_z \mu_B \left(m-m^\prime\right)\left(B-B_c\right)\right]^2},
\label{eq:splitting}
\end{equation}
where $B_c$ is the field of the center of the avoided crossing, $\Delta_{m,m^\prime}$ is the splitting at $B = B_c$, and $m$ and $m^\prime$ are the quantum numbers associated with the levels far from the avoided crossing --- 2 and -2, respectively, for the case studied here (for which $B_c=0$).
All energies are measured in units of frequency. The samples studied here were aligned to minimize $\theta$, the angle between the easy axis and the DC field, such that $\theta<10^{\circ}$, making Eq.~\ref{eq:splitting} a valid description of the splitting.

ESR measurements were done by placing single-crystal samples of $\nn$ in the loop of a loop-gap resonator (LGR)\cite{rinard_loop-gap_2005} designed to match specific frequencies. LGRs produce a uniform, high rf magnetic field in the loop, as shown schematically in Fig.~\ref{fig:lgrenfig}(b). The dimensions of the loop are small compared with the wavelength, allowing a high filling factor to be achieved.  Our LGRs typically have quality factors of $Q\sim2000$. Their resonant frequencies can also be tuned by introducing a dielectric such as sapphire into the gap, giving the resonators used in this work an effective range of several hundred MHz. It is also straightforward to fabricate new LGRs to have any desired frequency up through X band.  The experiments reported here were done using three separate LGRs.  Each LGR was placed inside a Cu shield (outer transparent cylinder in Fig.~\ref{fig:lgrenfig}(b)); the LGR is electrically isolated from the shield by nylon standoffs (solid grey).  A coaxial cable (brown) with a few-mm length of exposed inner conductor acts as an antenna, capacitively coupled to the LGR's gap.  We measure the reflected power ($S_{11}$) from the LGR using a vector network analyzer and determine the resonator's quality factor $Q$ from the response.  When the applied field brings the spin transition onto resonance with the LGR, the measured $Q$ drops.  While the results presented involved cw measurements,  LGRs can also be used with pulsed microwave excitations to study dynamics and determine relaxation times.  Since the transition studied in $\nn$ involves excited states, and therefore requires temperatures $\sim10$ K, relaxation was too fast to be measured with pulsed experiments.  However, the pulsed technique can be employed in other systems for which the ground-state tunnel splitting is on the order of the LGR's mode frequency.

\begin{figure}[tbh]
\includegraphics[width=1\linewidth]{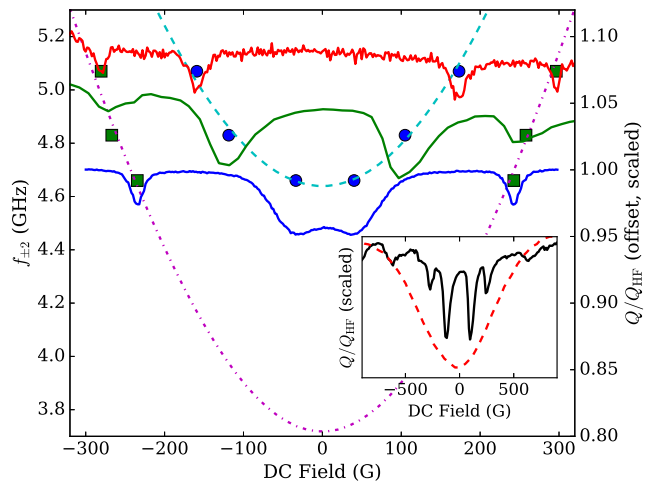}
\caption{\label{fig:ni4spectc}ESR spectra (right axis, $Q/Q_{\rm{HF}}$ vs.~$H$) for 4.655 (blue, lowest), 4.833 (green, middle), and 5.072~GHz (red, highest) at 10 K, offset by their frequency difference and scaled for clarity, where $Q_{\rm{HF}}$ is the high-field value of $Q$ for each trace. Transitions appear as dips in $Q/Q_{\rm{HF}}$. Blue circles and green squares show the frequency (left axis) and field locations of the transitions. The dashed blue and dash-dotted purple lines show fits of these data to Eq.~\ref{eq:splitting}, resulting in tunnel splittings of $\Do$~GHz and $\Dt$~GHz. Inset: Scaled spectra at 4.8~GHz and 10 K for 100\% $\nn$ (red dashed line) and dilute $(\nn)_{0.05}(\zz)_{0.95}$ (black solid line). The peaks at $\sim\pm600$~G in the dilute spectrum arise from impurities in the apparatus and are not associated with the sample.}
\end{figure}
Figure \ref{fig:ni4spectc} shows $(\nn)_{0.05}(\zz)_{0.95}$ (5\% dilute Ni$_4$) spectra taken at $f=4.655$, 4.833, and 5.072~GHz, showing peaks (symmetric around zero field) associated with transitions between the states shown in the inset of Fig.~\ref{fig:lgrenfig}; the peaks at lower (higher) fields correspond to transitions associated with the green solid (red dashed) levels.  For comparison purposes, the inset shows spectra at $f=4.8$~GHz for 100\% $\nn$ (red dashed line) and 5\% dilute $\nn$ (black solid line), illustrating the effectiveness of dilution in enabling the resolution of the fine features we are investigating (the four central peaks). 
The precision of our experiment relies on being able to resolve these fine features.  By fitting the spectral peaks for the dilute $(\nn)_{0.05}(\zz)_{0.95}$ sample to Lorentzian functions, we extracted the field location for each peak at each frequency. Figure \ref{fig:ni4spectc} shows the observed frequency-field relation for peaks from each conformational state as blue circles and green squares. We fit this data to Eq.~\ref{eq:splitting}; in determining $\Delta_{\pm2}$, the zero-field tunnel splitting, no assumptions need be made about $g_z$, which only affects the field dependence. We applied this fitting for both conformational states, and the resulting fits are shown as the blue dashed line and the purple dash-dotted line in Fig.~\ref{fig:ni4spectc}, yielding splitting values of $\Do$~GHz and $\Dt$~GHz. Using first-order perturbation theory, one can show that this splitting is related to the transverse anisotropy through $\Delta_{\pm2}=720C$, which gives $C_1=6.44(3)$ MHz and $C_2=5.16(27)$ MHz for the two conformational states. These values are in reasonably good agreement with previous measurements\cite{kirman_origin_2005,de_loubens_magnetization_2008} of $C=6$ MHz, but give much greater precision and allow us to differentiate the values associated with the two conformational states. The uncertainty in the determination of the splitting for the second state is significantly higher than the first, due to the lack of data near its zero-field frequency. Our measurement technique provides a determination of the transverse anisotropy parameter $C$ with unprecedented  (three-digit) precision, independent of the value of any other anisotropy parameters, thus avoiding the systematics that can arise from cross correlations among multiple fitting parameters. From the fit to Eq.~\ref{eq:splitting}, we also extract $g_z$ values of $g_{z,1}=2.18(8)$ and $g_{z,2}=2.11(16)$, which are consistent with each other and with values determined at much higher fields.\cite{chen_observation_2016}

Since our experiment involves applying the field along the easy axis of the system, we do not gain information about the direction of the hard axes (x and y) relative to the crystallographic directions.  A careful study of the behavior of the tunnel splitting on the azimuthal direction of a field applied in the x-y plane should allow a precise determination of the hard-axis directions.  In fact, when the field is applied along a hard axis of a four-fold symmetric SMM, a geometric-phase-interference effect should cause the tunnel splitting to be suppressed for certain field magnitudes,\cite{park_topological_2002,kim_quantum_2002,adams_geometric-phase_2013} similar to what has also been observed in two-fold symmetric molecules.\cite{wernsdorfer_quantum_1999-1,wernsdorfer_spin-parity_2002,wernsdorfer_quantum_2002}  For some SMMs, the geometric-phase interference has notably different predicted behavior for the giant-spin and more exact multispin models.\cite{barra_origin_2007,sorace_origin_2013}  So, an experimental study of this effect in $\nn$ may illuminate the relative validity of various models.

The symmetry of the molecule (S$_4$) allows us to attribute the measured tunnel splitting to Eq.~\ref{eq:hprime}, which is the leading-order term consistent with the symmetry. Without measurements of other tunnel splittings, we cannot rule out contributions from higher-order transverse-anisotropy terms in the Hamiltonian.  Neglecting the small effect of such possible terms, we can relate the splitting of the observed transition to $C$ through first-order perturbation theory.  It is worth noting that $C$ is a parameter within the so-called giant-spin approximation in which the system is treated as a single, large spin, a model that is only strictly valid in the limit of large intramolecular exchange interactions.  Interestingly, in $\nn$ a fourth-order transverse anisotropy cannot occur in that limit.\cite{yang_single-molecule_2005,wilson_magnetization_2006} Instead, mixing between different $S$ manifolds gives rise to the effective transverse anisotropy.  That said, for transitions restricted to the lowest-energy states, the effective giant-spin Hamiltonian appears to work remarkably well for $\nn$.\cite{chen_observation_2016} The technique described can also be applied to SMMs with lower symmetry, where the dominant transverse anisotropy would have the form $\frac{E}{2}\left(S_{+}^{2}+S_{-}^{2}\right)$ and the value of $E$ could be determined directly by measuring the splitting between $m=\pm1$ states at zero field.

\begin{acknowledgments}
We thank C.~Yoo for his work on the design and development of the resonators and J.~Kubasek for assistance in fabrication of the resonators.  We thank Dr.~R.J.~Cassella for performing metal analyses.  Support for this work was provided by the U.~S.~National Science Foundation under Grant No.~DMR-1310135 and by the Amherst College Dean of Faculty. R.A.A.C.~acknowledges CAPES and CNPq  for financial support. J.R.F.~acknowledges the support of the Amherst College Senior Sabbatical Fellowship Program, funded in part by the H.~Axel Schupf '57 Fund for Intellectual Life.
\end{acknowledgments}

\bibliography{spectrefs}{}

\end{document}